 \documentclass[final,5p,times,twocolumn]{elsarticle}
\usepackage{etoolbox}
\usepackage{lettrine}
\usepackage{braket} 
\usepackage{flushend} 
\usepackage{lipsum} 
\usepackage[final]{pdfpages} 

\patchcmd{\pprintMaketitle}{\begin{center}}{\begin{flushleft}}{}{}
\patchcmd{\pprintMaketitle}{\end{center}}{\end{flushleft}}{}{}
\patchcmd{\MaketitleBox}{\begin{center}}{\begin{flushleft}}{}{}
\patchcmd{\MaketitleBox}{\end{center}}{\end{flushleft}}{}{}

\usepackage{amssymb}
\usepackage{amsthm}
\usepackage{mathtools}

\usepackage[nodots,nocompress]{numcompress}

\usepackage{color}

\biboptions{super, compress}

\journal{arXiv}

\begin{document}

\begin{frontmatter}

\title{Predissociation of methyl cyanoformate: The HCN and HNC channels}

\author[temp]{Michael J. Wilhelm \corref{cor1}}
\ead{michael.wilhelm@temple.edu}
\author[sant]{Emilio Mart\'{i}nez-N\'{u}\~{n}ez \corref{cor1}}
\ead{emilio.nunez@usc.es}
\author[madr]{Jes\'{u}s Gonz\'{a}lez-V\'{a}zquez} 
\author[sant]{Saulo A. V\'{a}zquez} 
\author[temp]{Jonathan M. Smith} 
\author[temp]{\\ and Hai-Lung Dai}
\cortext[cor1]{Author to whom correspondence should be addressed:}
\address[temp]{Department of Chemistry, Temple University, 1901 N. $13^{th}$ St., Philadelphia, Pennsylvania 19122, USA.}
\address[sant]{Departamento de Qu\'{i}mica F\'{i}sica and Centro Singular de Investigaci\'{o}n en Qu\'{i}mica Biol\'{o}xica e Materiais Moleculares (CIQUS), Universidade de Santiago de Compostela, 15782, Santiago de Compostela, Spain.}
\address[madr]{Departamento de Qu\'{i}mica, Universidad Aut\'{o}noma de Madrid, M\'{o}dulo 13, 28049 Madrid, Spain.}

\begin{abstract}
We present a combined experimental and theoretical investigation of the 193 nm photolysis of the cyano-ester, methyl cyanoformate (MCF). Specifically, nanosecond time-resolved infrared emission spectroscopy was used to monitor the ro-vibrationally excited photoproducts generated in the photolysis reaction. The signal-to-noise of all time-resolved spectra were enhanced using the recently developed algorithm, spectral reconstruction analysis, which allowed observation of previously obscured minor resonances, and revealed new dissociation channels producing HCN and HNC. Spectral fit analysis of the nascent HCN and electronically excited CN($A^2\Pi_1$) resonances yield a lower bound estimate for the HCN quantum yield of ca. 0.42$\pm$0.24\%. Multi-configuration self-consistent field calculations were used to interrogate the ground and four lowest energy singlet excited states of MCF. At 193 nm, dissociation is predicted to occur predominantly on the repulsive S$_2$ state. Nevertheless, minor pathways leading to the production of highly excited ground state MCF are available via cascading internal conversion from nascent S$_2$ population. An automated transition-state search algorithm was employed to identify the corresponding ground state dissociation channels, and Rice-Ramsperger-Kassel-Marcus and Kinetic Monte Carlo kinetic simulations were used to calculate the associated branching ratios. The proposed mechanism was validated by direct comparison of the experimentally measured and quasi-classical trajectory deduced nascent internal energy distribution of HCN, which were found to be in near perfect agreement. The propensity for cyano containing hydrocarbons to act as photolytic sources for ro-vibrationally excited HCN and HNC, as well as their significance to astrophysical environments, are discussed.
\end{abstract}


\end{frontmatter}


\noindent
\lettrine[lines=2]{H}{ydrogen} cyanide (HCN) and hydrogen isocyanide (HNC) are well known constituents of the interstellar medium (ISM){\color{blue}\cite{Barger2003, Bechtel2006, Blackman1976, Brown1977, Brown1976, Gao2007, Hirota1998, Irvine1998, Petrie2001, Rodgers1998, Schilke2003, Ziurys1999}}. Both species have been observed in a variety of astrophysical environments, including: comets{\color{blue}\cite{Irvine1998, Rodgers1998, Irvine1996, Irvine1998a}}, dark clouds{\color{blue}\cite{Hirota1998}}, and proto-planetary nebulae{\color{blue}\cite{Schilke2003}}. Of significance, despite the fact that HNC is roughly 14.3 kcal/mol less stable than HCN (i.e., suggesting an infinitesimal 100 K equilibrium HNC/HCN abundance ratio of ca. 10$^{-30}$), HNC is often observed in comparable concentrations to HCN{\color{blue}\cite{Brown1977}}. The origin of this discrepancy is a long standing puzzle in the astrophysical community{\color{blue}\cite{Barger2003, Bechtel2006, Hirota1998, Irvine1998, Rodgers1998, Ziurys1999, Irvine1996}}. Following detection of HNC in comet C/1996 B2 (Hyakutake), Irvine et al. outlined a series of mechanisms to account for the unexpectedly large abundance of the isocyanide, including UV photolysis of larger parent cyanides{\color{blue}\cite{Irvine1996}}. In principle, all cyano containing hydrocarbons are potential photolytic sources of HCN and HNC. It is therefore of interest to perform a systematic investigation of the UV photodissociation dynamics of each of the distinct representative classes of cyano containing hydrocarbons.\\
\indent We have previously characterized the 193 nm photolysis of vinyl cyanide (H$_2$C=CHCN), and revealed it to be a viable photolytic source of both HCN and HNC, with a relative HNC/HCN abundance ratio of ca. 0.3{\color{blue}\cite{Wilhelm2009, Homayoon2011, Vazquez2015}}. It must be noted that the prior photofragment translational energy spectroscopy (PTS) study of this reaction correctly identified HCN as a photofragment, but made no mention of the relative presence of HNC{\color{blue}\cite{Blank1998b}}. This deficiency stemmed from the fact that the PTS detection scheme relied primarily on the mass of the photofragments{\color{blue}\cite{Blank1998b}}. Ultimately, when mass alone is used to identify photofragments, geometric isomers are easily overlooked. Indeed, in their investigation of the UV photolysis of the diazine heterocycle, pyrimidine (C$_4$N$_2$H$_4$), Lee and colleagues acknowledged the production of a m/e=27 photofragment, but were unable to conclude whether it originated from HCN, HNC, or both{\color{blue}\cite{Lin2006}}. Subsequently, in order to definitively identify the presence of geometric isomers in photolysis reactions, it is crucial to employ a detection scheme sensitive to molecular structure (e.g., infrared emission).\\
\indent We now examine the UV photolysis of the cyano-ester, methyl cyanoformate (NCC(O)OCH$_3$, MCF). Furlan et al. previously characterized the 193 nm photolysis of MCF using PTS, in which they focused exclusively on the Norrish type I dissociation channels{\color{blue}\cite{Furlan2000}}. They measured photofragment time-of-flight signals suggestive of NCCO, CH$_3$OCO, CN, CO, and CH$_3$.  It was concluded that the photolysis resulted in two primary dissociation channels:
\begin{equation}
NCC(O)OCH_3 \rightarrow NCCO + CH_3O; \Phi=95\pm2\%
\end{equation}
\begin{equation}
NCC(O)OCH_3 \rightarrow CN + CH_3OCO; \Phi=5\pm2\%
\end{equation}
\noindent Later, in a time-resolved infrared diode laser (IDL) study aimed at monitoring the reaction kinetics of the cyanooxymethyl (NCCO) radical, Feng and Hershberger described a third channel involving unimolecular dissociation of internally excited NCCO$^{\ddagger}$ produced in reaction (1){\color{blue}\cite{Feng2010}}:\\
\begin{equation}
NCCO^{\ddagger}  \rightarrow CN + CO; \Phi=38\pm0.09\%
\end{equation}
\noindent Nevertheless, neither the PTS nor the IDL study reported observation of either HCN or HNC.\\
\indent Time-resolved IR emission spectra following the 193 nm photolysis of MCF have previously been reported{\color{blue}\cite{McNavage2004}}. Specifically, in an effort to identify the fundamental vibrational modes of the NCCO radical, McNavage et al. used time-resolved Fourier transform infrared emission spectroscopy (TR-FTIRES) to interrogate the photolysis of MCF{\color{blue}\cite{McNavage2004}}. As distinct from PTS, in which different reaction channels are individually examined in separate experiments, TR-FTIRES yields spectra composed of simultaneous contributions from all vibrationally excited photofragments. Subsequently, a core benefit of TR-FTIRES for characterizing unimolecular dissociation reactions is that unexpected dissociation channels are more readily detected. Nevertheless, while complete time-resolved IR emission spectra of MCF were collected (ca. 700-4000 cm$^{-1}$ spectral range), they were convoluted into a two-dimensional cross-correlation analysis with photolysis spectra from other NCCO precursors{\color{blue}\cite{McNavage2004}}. Beyond an assessment of the relative contribution of the NCCO radical, the IR emission spectra collected following the 193 nm photolysis of MCF have yet to be documented.\\
\indent In this report, we present a combined experimental and theoretical examination of the UV photodissociation dynamics of MCF, with a specific focus on channels leading to the production of HCN or HNC. Specifically, we characterize the time-resolved IR emission spectra collected following the 193 nm photolysis of MCF. In an effort to tease out comparatively weak signals, a recently developed algorithm for enhancing the signal-to-noise ratio (SNR) in time-resolved spectroscopies was universally applied to all spectra{\color{blue}\cite{Wilhelm2015d}}. Additionally, multi-configuration self-consistent field calculations were used to calculate potential energy curves for the ground and four lowest energy singlet excited states of MCF, as a means of investigating the dominant NCCO + CH$_3$O dissociation channel (reaction 1). Further, a recently developed automated transition state (TS) search algorithm was used to search for HCN and HNC elimination channels in the ground electronic state of MCF{\color{blue}\cite{Martinez-Nunez2015, Martinez-Nunez2015a}}. Branching ratios for all identified dissociation channels were deduced using a combination of Rice-Ramsperger-Kassel-Marcus (RRKM) theory{\color{blue}\cite{Smith1990}} and Kinetic Monte Carlo (KMC) simulations{\color{blue}\cite{Gillespie1976}}.\\
\\
{\bf{METHODOLOGY}}
\subsection{\bf{\em{Time-resolved Fourier transform IR emission spectroscopy.}}}
Time-resolved IR emission spectroscopy allows characterization of the temporal evolution of excited molecules, and is a powerful tool for monitoring photolysis reactions{\color{blue}\cite{Wilhelm2009, Hancock2005, Heazlewood2008, Liu2010, Yeh2012, Wilhelm2013a, Ma2012}}. It must be stressed, however, that emission spectroscopy measures signal solely from excited species. Subsequently, any photofragment that is produced vibrationally cold will not be recorded in the IR spectra. Conversely, provided the photofragments exhibit sufficiently strong transition dipole moments, even relatively minor elimination channels can be detected.\\
\indent The setup for the TR-FTIRES experiment has been described in detail previously{\color{blue}\cite{Wilhelm2013a, Hartland1992}}. Briefly, the 193 nm output from an ArF excimer laser (Lambda Physik, EMG 201 MSC, 20 Hz, $\leq$30 mJ pulse$^{-1}$) was collimated and lightly focused through a photolysis cell mounted with two CaF$_2$ salt windows. Pressure within the cell was maintained with 30 mTorr of the MCF precursor gas, and 4 Torr of an inert Ar buffer gas under constant flow conditions. Total pressure was monitored with a capacitance manometer (MKS Baratron, 0-10 Torr). Emission after the photolysis pulse was collected perpendicular to the laser propagation axis by a set of gold-plated mirrors arranged as a Welsh cell, then collimated and focused into the FTIR spectrometer (Bruker IFS-88) with two KBr salt lenses. The spectrometer was equipped with an interferometer capable of time-resolved step-scan measurements and a liquid nitrogen cooled mercury-cadmium-telluride (MCT) IR detector (Judson Technologies, EG$\&$G, J15D14).\\
\indent The MCT detector has a highly peaked frequency sensitivity, exhibiting a 100\% maximum near 900 cm$^{-1}$, which drops to ca. 30\% around 1500-2500 cm$^{-1}$, and $<$15\% above 3000 cm$^{-1}$. The frequency response of the system was calibrated using the mid-IR Globar in the spectrometer, which was modeled as a 1300 K blackbody source.  Interferograms were recorded in 0.05 $\mu$s steps over a 20 $\mu$s interval in which signal acquisition was triggered by a reflection of the photolysis laser pulse off the front window of the reaction cell. Each interferogram point was obtained by averaging the signal from 64 photolysis events. The interferometric signal was amplified ten times before reaching the transient digitizer (Tektronix, TDS380). Fourier transformation of the interferograms yields time-resolved spectra in increments of 0.05 $\mu$s following the arrival of the photolysis pulse. Spectral resolution was fixed at 12 cm$^{-1}$.\\
\indent Prior to use, commercially available MCF (Aldrich, 99\%) was purified by several freeze-pump-thaw cycles. Argon gas (Spectra Gas, research grade, 99.9\%) was used directly as obtained from the supplier.
\subsection{\bf{\em{Post-processing signal analysis.}}}
A new algorithm for enhancing the SNR in time-resolved spectroscopies, termed spectral reconstruction analysis (SRa), has recently been described{\color{blue}\cite{Wilhelm2015d}}. Specifically, for a series of {\em{n}} time-resolved spectra, SRa yields an upper bound enhancement  of ca. 0.6$\sqrt{n}$ in the SNR of each of the {\em{n}} spectra analyzed{\color{blue}\cite{Wilhelm2015d}}. As distinct from a simple linear average, which produces only a single representative spectrum with enhanced SNR, SRa improves the SNR for all spectra, and therefore fully preserves the measured temporal dependence.\\
\indent Briefly, SRa operates by eliminating noise in the time-domain, thereby significantly attenuating noise in the frequency-domain. Specifically, the time-dependence at each measured frequency is fit to a mathematical function to capture the general temporal evolution. Reconstructed spectra are then generated by replacing the set of measured time-dependence curves with the resulting noiseless fit curves.\\
\indent For the current study, SRa of the time-resolved MCF photolysis spectra involved individually fitting the temporal-dependence of all the measured frequency components to a general sum of three rising and/or decaying exponential functions:
\begin{equation}
f(t) = \phi_0 + \sum_{i=1}^{3}{(-1)^{i+1}\phi_i\times exp(\pm k_it)},
\end{equation}
where $\phi_0$ is a baseline offset, $\phi_{i\geq 1}$ are intensity scaling factors, and k$_i$ are rate constants. SRa was automated in {\sc{Matlab}} (Mathworks), where the fit analysis employed the {\sc{CreateOptimProblem}} structure within {\sc{Matlab}}'s {\sc{LsqCurveFit}} solver function.
\subsection{\bf{\em{Theoretical calculations of the dissociation channels.}}}
The NCCO + CH$_3$O dissociation channel was investigated by multiconfigurational self-consistent field calculations, using the cc-pVTZ correlation consistent basis set{\color{blue}\cite{Dunning1989}}. Potential energy curves, as a function of the C-O distance, were computed for the five lowest lying singlet states. For all the curves, the calculated points correspond to ground state optimized geometries obtained by state average CASSCF{\color{blue}\cite{Roos1980}} calculations over the five roots equally weighted (SA5). The active space consisted of 12 electrons in 10 orbitals, as described in the {{\em{Supplementary Information}} (SI). The energies were computed by single-point multi-state CASPT2{\color{blue}\cite{Roos1987}} (MSCASPT2) calculations, using the previously calculated SA5-CASSCF(12,10) wave functions as references. Full geometry optimizations, relaxing the appropriate root, were also carried out to calculate the equilibrium structure for each state. The excited state calculations were performed with the {\sc{Molcas 7}} program{\color{blue}\cite{Veryazov2004, Aquilante2010}}.\\
\indent The UV spectrum of MCF was simulated by using a set of 100 initial uncorrelated geometries and velocities generated by a Wigner harmonic distribution of the vibronic ground state using MP2/6-311+G(2d,2p) frequencies. A single point calculation employing the same previously described protocol (SA5-CASSCF(12,10)/cc-pVTZ corrected with MS5-CASPT2) was performed for every initial condition and the energy and dipole moment were extracted using the Perturbed Modified CAS wavefunction as implemented in {\sc{Molcas}}{\color{blue}\cite{Veryazov2004, Aquilante2010}}. Finally, the spectrum was created by using a Gaussian envelope with a width of 2.3 kcal/mol in every trajectory.\\
\indent To investigate dissociation channels in the ground electronic state (S$_0$), and in particular to find possible HCN elimination mechanisms, the automated transition state (TS) search algorithm TSSCDS, developed by one of the authors, has been employed{\color{blue}\cite{Martinez-Nunez2015, Martinez-Nunez2015a}}. A detailed description of the method and the parameters employed in this work are given in the SI.\\
\indent Two different levels of theory were employed in the TSSCDS procedure. The latest Stewart's PM7 semiempirical method{\color{blue}\cite{Stewart2013}} was used for the exploratory dynamics simulations, as well as for an initial optimization of the TSs using {\sc{Mopac2012}}{\color{blue}\cite{Stewart2012}}. Then, the structures were reoptimized at the B3LYP/6-31+G(d,p) and MP2/6-311+G(2d,2p) levels of theory. For the energies, single point CCSD(T)/6-311+G(3df,2p) calculations were employed (see the SI for further details). The DFT and coupled cluster calculations were performed with {\sc{Gaussian09}}{\color{blue}\cite{Frisch2009}}.\\
\indent The branching ratios of the dissociation channels of MCF in its S$_0$ state were computed using a combination of RRKM theory{\color{blue}\cite{Smith1990}} and KMC{\color{blue}\cite{Gillespie1976}} simulations. A full description of these calculations is given in the SI. Briefly, RRKM rate coefficients for each process were computed for an excitation energy of 148.12 kcal/mol, which corresponds to a photon wavelength of 193 nm. KMC simulations were then employed to follow the transient population of MCF, intermediates, and products. The kinetic analysis provides branching ratios for the various dissociation channels found in this study in the S$_0$ state.\\
\\
\noindent {\bf{\em{Quasi-classical trajectory (QCT) simulations.}}}\vspace{0.5em}\\
HCN product energy distributions were calculated in this work using QCT simulations. The PM3 Hamiltonian{\color{blue}\cite{Stewart1989}} with specific reaction parameters (SRP){\color{blue}\cite{Gonzalez-Lafont1991}} was employed to evaluate energies and gradients. A description of the PM3 reparametrization is gathered in the SI.
\indent The general chemical dynamics programs {\sc{Venus05}}{\color{blue}\cite{Hase1996}} was employed to carry out the simulations. Since the aim was to simulate the HCN elimination pathway observed in the 193 nm (148 kcal/mol) photolysis of the molecule, an excess vibrational energy of 72.7 kcal/mol is placed at the transition state leading to HCN + CO + CH$_2$O (TS37; see Figure 7 below) using quasi-classical normal mode sampling{\color{blue}\cite{Bolton}} and zero rotational energy. An ensemble of around 50000 trajectories was integrated using a sixth-order Adams Moulton predictor-corrector algorithm with a fixed time step of 0.1 fs, which ensured an average energy conservation of 99.99\%.\\
\indent The integration finishes when all inter-fragment distances reached at least 15 $\AA$. The rotational quantum number J of HCN was obtained by equating the modulus of the classical rotational angular momentum to [J(J + 1)]$^{1/2}\hbar$. The vibrational states of hydrogen cyanide ($\nu_1$,$\nu_2^l$,$\nu_3$), where $\nu_1$, $\nu_2$, $\nu_3$ and {\em{l}} refer to the CH stretch, bend, CN stretch, and vibrational angular momentum, respectively, were also calculated here using the normal mode analysis approach of Corchado and Espinosa-Garcia{\color{blue}\cite{Corchado2009}}. In this approach, the vibrational angular momentum {\em{l}} of HCN is computed from j$_a$  = {\em{l}}$\hbar$, where j$_a$ is the projection of the rotational angular momentum of the molecule j onto the molecular axis. The (real) values of the classical actions thus obtained are rounded to the nearest integers to obtain vibrational quantum numbers. The probabilities of the vibrational states were obtained using both the standard histogram binning (HB), and an energy-based Gaussian binning (1GB) procedure{\color{blue}\cite{Czabo2009, Czabo2012, Bonnet1997, Bonnet2004}} (more details are given in the SI).
\section{RESULTS and ANALYSIS}
\subsection{\bf{\em{Time-resolved IR emission spectra following 193 nm photolysis of MCF.}}}
Figure 1 depicts representative time-resolved IR emission spectra (spanning 700-3949 cm$^{-1}$), including both unprocessed measured spectra (left) and SRa results (right), which were collected following the 193 nm photolysis of MCF at 0.5 (grey),
\begin{figure}[h]
 \centering
  \includegraphics[height=6.5cm]{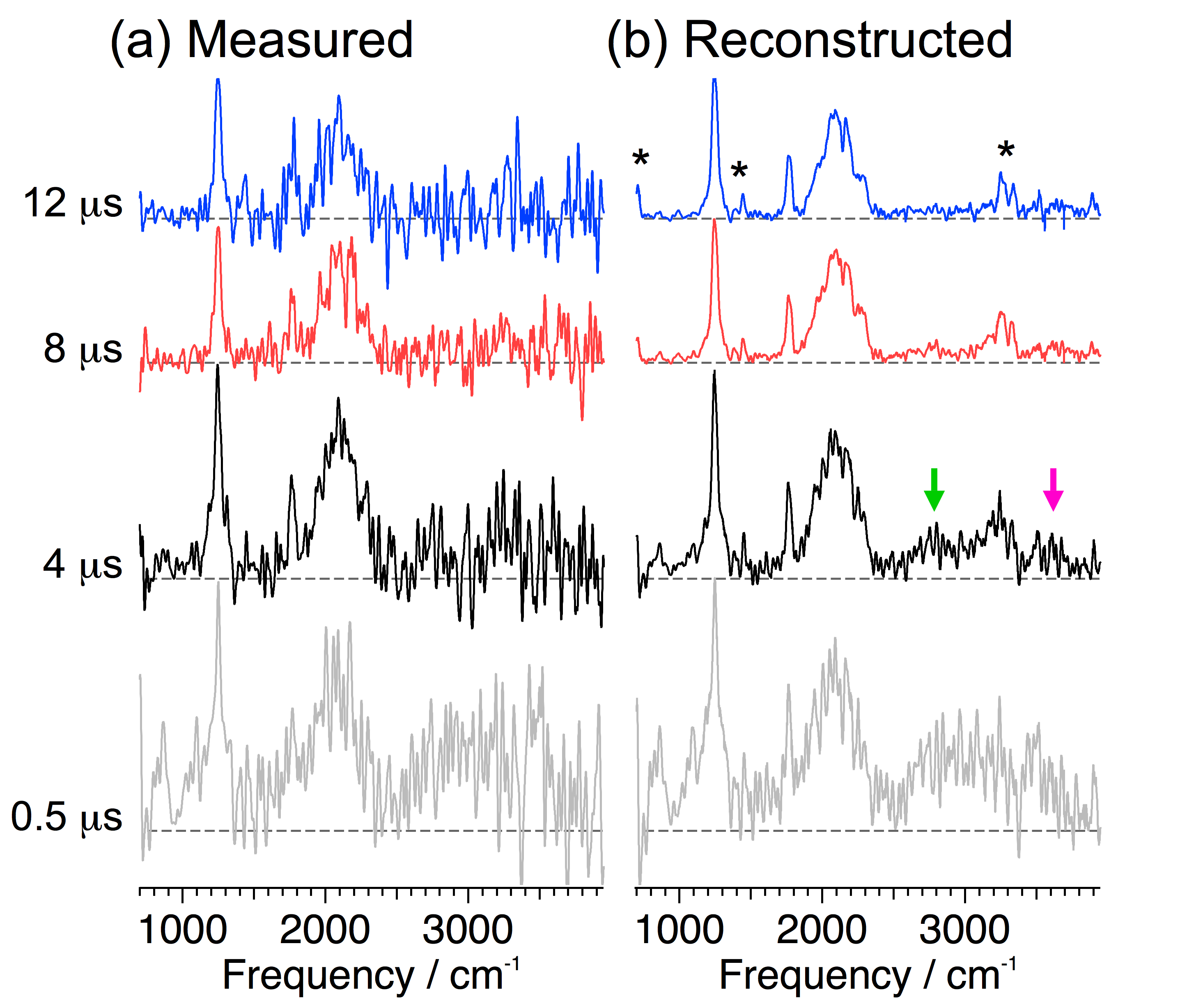}
  \caption{Representative time-resolved (a) measured and  (b) reconstructed IR emission spectra collected following the 193 nm photolysis of MCF at 0.5 (grey), 4 (black), 8 (red), and 12 (blue) $\mu$s, following the arrival of the photolysis pulse. Observation of new resonances in the reconstructed spectra are indicated with asterisks and colored arrows.}
\end{figure}\\
4 (black), 8 (red), and 12 $\mu$s (blue). As demonstrated previously{\color{blue}\cite{Wilhelm2015d}}, SRa increases the SNR of each of the time-resolved spectra (ca. 5x improvement), resulting in better resolved resonances. Significantly, due to the reduction of noise, the reconstructed spectra show additional features which had previously been obscured in the measured spectra. Specifically, the reconstructed spectra show evidence of five new resonances not present in the measured spectra: a transient band centered near 2785 cm$^{-1}$ (highlighted with a green arrow in the 4 $\mu$s spectrum), a weak band near 3650 cm$^{-1}$ (highlighted with a pink arrow in the 4 $\mu$s spectrum),and three stable resonances (indicated with an * in the late spectrum) near 700 cm$^{-1}$, 1410 cm$^{-1}$, and 3310 cm$^{-1}$.\\
\begin{figure}[h]
 \centering
  \includegraphics[height=6.5cm]{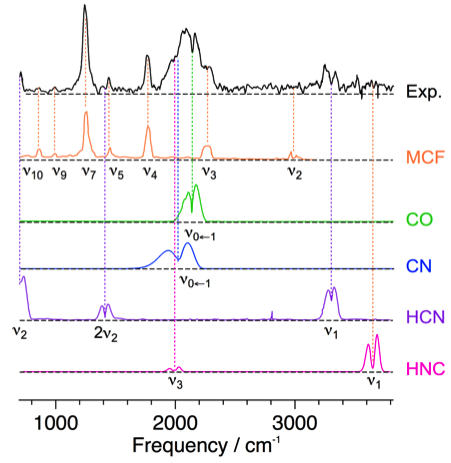}
  \caption{Comparison of a late time (12 $\mu$s) reconstructed IR emission spectrum (black) collected following the 193 nm photolysis of MCF, and room-temperature IR absorption spectra of MCF (orange), CO (green), CN (blue), HCN (purple), and HNC (pink). Associated fundamental vibrational modes have been identified beneath each spectrum.}
\end{figure}
\indent Figure 2 depicts spectral assignments for a representative late time (12 $\mu$s) SRa enhanced spectrum. By this time, the photofragments have experienced ca. 500 thermalizing collisions with Ar, and therefore originate predominantly from fundamental transitions. Specifically, in addition to photofragments, the measured spectra show clear evidence of vibrationally excited MCF precursor (orange trace), which can be produced either by energy transferring collisions, or from re-association of energized photofragments. Beyond the precursor emission bands, the most prominent feature is a heavily convoluted band spanning ca. 1800-2350 cm$^{-1}$. The low energy side of this band may contain contributions from the $\nu_2$ CO stretch of NCCO (1889 cm$^{-1}$ fundamental{\color{blue}\cite{Feng2010, Feng2011}}). Nevertheless, due to the spectral congestion of this band, any contribution from NCCO is completely unresolvable. Overall, the majority of the emission intensity can be assigned to carbon monoxide (CO, green trace), the cyano radical (CN, blue trace), and the $\nu_3$ CN stretch of MCF. Conversely, as highlighted in Figure 1, the fast decay kinetics of the new resonance near 2785 cm$^{-1}$ suggests an origin from a primary transient photofragment, and can be reasonably assigned to emission from the $\nu_4$ CH$_3$ stretch of the methoxy radical (CH$_3$O){\color{blue}\cite{Chiang1989}}.\\
\indent It is important to note that the remainder of the newly detected resonances cannot be assigned to the MCF precursor, nor to any of the previously observed Norrish type I photofragments. The frequency range of the ca. 3310 cm$^{-1}$ resonance suggests a C-H stretch from an sp-hybridized carbon. Further, the partially resolved rotational contours of the 1410 and 3310 cm$^{-1}$ resonances, which exhibit well-defined P and R branches, strongly suggests the emission originates from a linear molecule. Altogether, the three new resonances can be assigned to emission from HCN (Figure 2, purple trace); specifically, the $\nu_1$ CH stretch (3311 cm$^{-1}$ fundamental), the $2\nu_2$ hot-band (1410 cm$^{-1}$ fundamental), and the $\nu_2$ bend (712 cm$^{-1}$ fundamental), respectively. Note that while the $\nu_3$ CN stretch of HCN at 2097 cm$^{-1}$ is an IR allowed transition, the transition dipole moment is prohibitively weak (i.e., $|\mu_{\nu^{'}}|^2=2\times 10^{-6} D^2$, ca. 4000 times weaker than the $\nu_1$ CH stretch), and hence would not be observable in the collected spectra. Additionally, the weak resonance near 3650 cm$^{-1}$ is tentatively assigned as originating from the $\nu_1$ NH stretch of HNC (Figure 2, pink trace). With respect to the frequency range of our MCT detector, the $\nu_1$ NH stretch (3652 cm$^{-1}$ fundamental) is the strongest detectable mode of HNC. Though, the $\nu_3$ CN stretch (2023 cm$^{-1}$ fundamental) may be contributing minor intensity to the convoluted band spanning ca. 1800-2350 cm$^{-1}$. Note that, for equal concentrations, the $\nu_1$ NH stretch of HNC would be roughly three times more intense than the $\nu_1$ CH stretch of HCN. Subsequently, the comparatively weak intensity of the observed HNC resonance suggests an HNC/HCN abundance ratio $<$ 1.\\
\indent Further, as part of a search for electronically excited photofragments, higher energy emission spectra, sampling out to 7898.5 cm$^{-1}$, were also measured. Figure 3 depicts representative (a) measured and (b) SRa treated spectra 0.5 $\mu$s following the arrival of the photolysis pulse.
\begin{figure}[h]
 \centering
  \includegraphics[height=6.5cm]{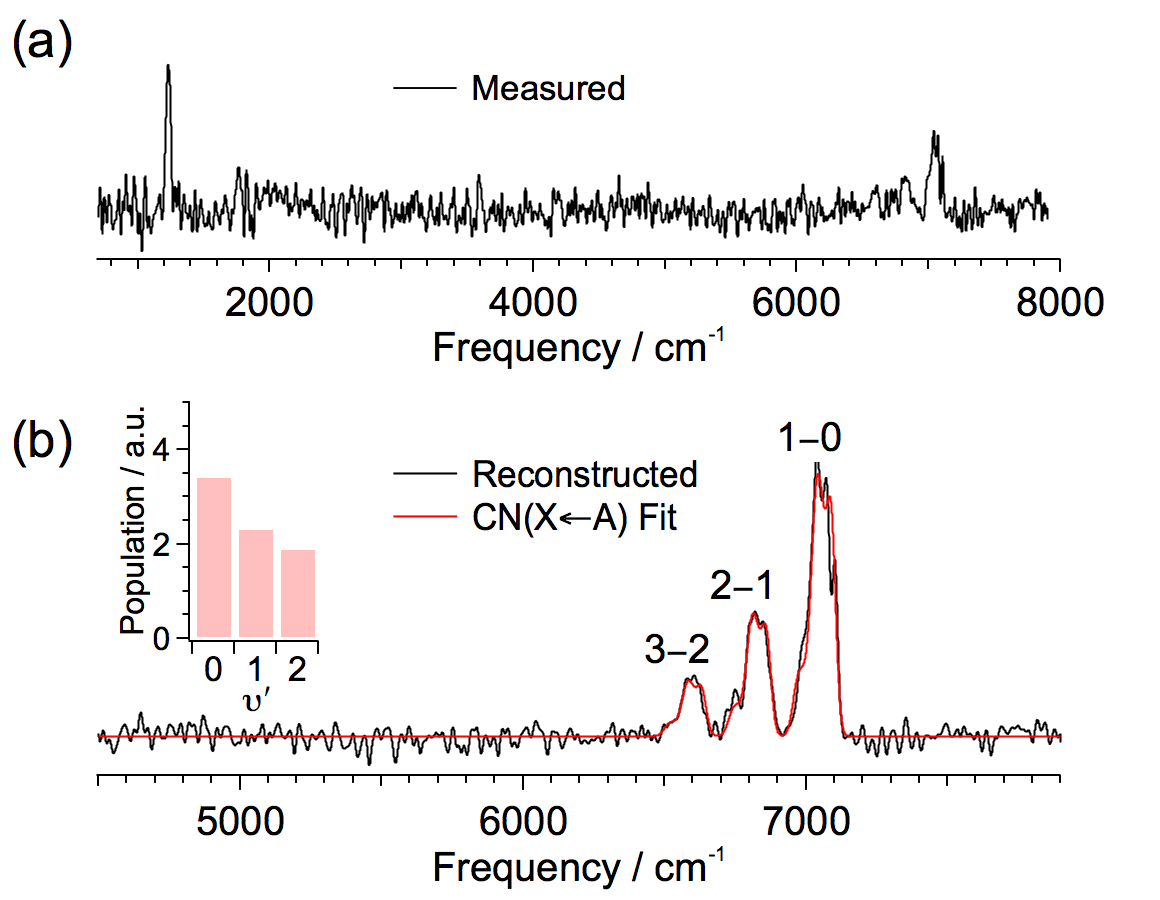}
  \caption{Early time (0.5 $\mu$s) (a) measured and (b) SRa enhanced high-energy IR emission spectrum showing vibronic emission from CN(A$^2\Pi_1$). A representative best fit is shown as an overlaid red trace. Inset depicts the fit-deduced vibrational population.}
\end{figure}
As depicted in the SRa enhanced spectrum (Figure 3b), three distinct ro-vibronic transitions are clearly visible at 6595 cm$^{-1}$, 6826 cm$^{-1}$, and 7051 cm$^{-1}$. These correspond to the so-called CN($X^2\Sigma^{+}\leftarrow A^2\Pi_1$) red system{\color{blue}\cite{Weinberg1967}}, and are assignable to the X($\nu^{"}$)$\leftarrow$A($\nu^{'}$) transitions: 3$\leftarrow$2, 2$\leftarrow$1, and 1$\leftarrow$0, respectively.
\subsection{\bf{\em{Photodissociation product population distributions.}}}
Emission spectra are sensitive indicators of internal energy, where, due to the anharmonic nature of the vibrational potential, transitions from highly excited levels are red-shifted relative to the fundamental.  Subsequently, by performing a spectral fit analysis, it is possible to extract the time-dependent population distribution as a function of the internal energy. Of significance for the current study, analysis of the CN ro-vibronic and HCN vibrational emission allows determination of the relative branching ratio for these channels.
\subsubsection{\bf{\em{1. Modeling ro-vibronic transitions of CN($X^2\Sigma^{+}\leftarrow A^2\Pi_1$).}}}
In order to measure the CN population distribution, it is necessary to perform a fit analysis of the measured spectrum (Figure 3b). Individual ro-vibronic transitions of the CN($X^2\Sigma^{+}\leftarrow A^2\Pi_1$)  red system were calculated using the rigid-rotor / harmonic oscillator selection rules (i.e., $\Delta\nu$=+1, $\Delta$J=0,$\pm$1), in which emission intensity was modeled as:
\begin{equation}
I_{\nu',J'}^{X\leftarrow A} = \omega_{\nu',J'}^3\braket{\nu''|\nu'}|\mu_{\nu'',\nu'}|^2S_{J'}^{\Delta J}\times exp\left(\frac{-E_{\nu',J'}}{k_BT_{rot}}\right),
\end{equation}
where $\omega_{\nu',J'}$  is the transition frequency, $\braket{\nu''|\nu'}$ is the Franck-Condon factor, $|\mu_{\nu'',\nu'}|$ is the transition dipole moment (i.e., deduced as the harmonically-scaled transition dipole moment of the fundamental transition, $\nu'|\mu_{0,1}|$), $S_{J'}^{\Delta J}$ is the H\"{o}nl-London factor, $E_{\nu',J'}$ is the energy of the emitting level, and T$_{rot}$ is the rotational temperature. Based upon the observed emission features (Figure 3b), transitions were calculated for upper state vibronic levels corresponding to $\nu'$=0,1, and 2. Representative basis spectra, $S_{\nu'}(\omega)$, for each of these three vibronic levels were individually constructed by summing over all allowed ro-vibronic transitions originating from the level ($\nu',J'$):
\begin{equation}
S_{\nu'}(\omega) = \sum_{J'}{I_{\nu',J'}^{X\leftarrow A}\times g\left(\omega_{\nu',J'},\sigma\right)},
\end{equation}
\noindent where $g\left(\omega_{\nu',J'},\sigma\right)$ is a Gaussian function centered at the transition frequency, $\omega_{\nu',J'}$ , with full-width at half maximum (fwhm), $\sigma$, reflecting the experimentally measured spectral resolution (ca. 12 cm$^{-1}$).\\
\indent In order to determine the total population of nascent CN($A^2\Pi_1$), the early time SRa treated spectrum in Figure 3b was fit using the simulated basis spectra constructed above. As there are only three observable vibronic transitions, the analysis allowed the population of each vibronic level to vary independently to obtain the best fit:
\begin{equation}
fit(\omega) = \phi + \Gamma_{CN}\sum_{i=0}^{2}{n_iS_i(\omega)},
\end{equation}
\noindent where $\phi$ is a baseline offset, $n_i$ is the population of the i$^{th}$ emitting level, and $\Gamma_{CN}$ is a global intensity scaling factor. As depicted in the inset of Figure 3b, the best fit result corresponds to a normalized vibrational population distribution of: ($\nu'$=0:$\nu'$=1:$\nu'$=2)$\rightarrow$(1.0:0.68:0.56). For the detection system employed, this suggests a CN($A^2\Pi_1$) population ($\Gamma_{CN}\Sigma_{i=0}^{2}n_i$) of ca. 7.7$\times$10$^{-7}$.
\subsubsection{\bf{\em{2. Modeling ro-vibrational transitions of HCN($\tilde{X}^1\Sigma^+$).}}}
Similarly, individual ro-vibrational transitions for the $\nu_1$ and $\nu_2$ (i.e., $\Delta\nu$=-1, $\Delta$J=0,$\pm$1) and $2\nu_2$ (i.e., $\Delta\nu$=-2, $\Delta$J=0,$\pm$1) modes of HCN($\tilde{X}^1\Sigma^+$) were calculated, in which emission intensity was modeled as:
\begin{equation}
I_{\nu',J'} = \omega_{\nu',J'}^3|\mu_{\nu'',\nu'}|^2S_{J'}^{\Delta J}\times exp\left(\frac{-E_{\nu',J'}}{k_BT_{rot}}\right),
\end{equation}
\noindent where the upper state level, $\nu'$, corresponds to the set of vibrational quantum numbers: ($\nu_1,\nu_{2}^{l},\nu_3$). Similar to the CN vibronic spectral basis set, each HCN ro-vibrational transition was broadened by a Gaussian function whose width matched the 12 cm$^{-1}$ spectral resolution of the experiment. Unlike diatomic CN, due to the density of states of HCN, a spectral fit analysis using a basis set composed of all individual vibrational transitions is computationally intractable. Rather, the ro-vibrational manifold of HCN was partitioned into discrete energy bins, each 500 cm$^{-1}$ in width. As described previously{\color{blue} \cite{Wilhelm2009, Ma2012, Hartland2001, Nikow2010, Smith2014}}, all transitions originating from a given bin were then summed together to produce a single representative bin spectrum, in which all levels within the bin were assumed to be equally populated. However, as internal energy increases, multiple transitions are possible from a single common upper state (e.g., [$\nu_1,\nu_{2}^{l},\nu_3$]  $\rightarrow$ [$\nu_1-1,\nu_{2}^{l},\nu_3$], [$\nu_1,\nu_{2}^{l\pm0,1}-1,\nu_3$], or [$\nu_1,\nu_{2}^{l}-2,\nu_3$]). Subsequently, each spectrum within a bin was inversely weighted by the total number of transitions from the associated upper state, as a pseudo degeneracy factor. Overall, HCN emission was modeled using a total of (23) basis spectra, spanning the zero point energy (ca. 3412.5 cm$^{-1}$) up to 15000 cm$^{-1}$. The time-dependent population of each bin, N$_i$(E$_i,\gamma$), was modeled as:
\begin{equation}
N_i(E_i,\gamma) = \rho_i\left(\gamma\sqrt{2\pi}\right)^{-1}exp\left(\frac{-E_{i}^{2}}{2\gamma^2}\right),
\end{equation}
\noindent where E$_i$ is the energy of bin i, $\rho_i$ is the number of pure vibrational states which contribute to emission from bin i, and $\gamma$ is the width of the total distribution. Overall, as shown in Figure 4, the HCN emission was fit by varying the width of the distribution, $\gamma$:
\begin{equation}
fit(\omega) = \phi + \Gamma_{HCN}\sum_{i=1}^{23}{N_i(E_i,\gamma)\times S_i(\omega)},
\end{equation}
\noindent where $\phi$ is a baseline offset, the S$_i(\omega)$ are the HCN bin spectra, and $\Gamma_{HCN}$ is a global intensity scaling factor. Note that the extra intensity in the 2$\nu_2$ stretch near 1410 cm$^{-1}$, which is not reproduced in the fit analysis, stems from spectral overlap of the $\nu_5$ mode of MCF. Analysis of the nascent HCN emission yields a population fwhm of ca. 6000 cm$^{-1}$, the tail of which extends out beyond 15000 cm$^{-1}$. For the detection system employed, this corresponds to an HCN($\tilde{X}^1\Sigma^+$) population ($\Gamma_{HCN}\Sigma_{i=1}^{23}N_i(E_i,\gamma)$) of ca. 6.4$\times$10$^{-8}$.
\begin{figure}[h]
 \centering
  \includegraphics[height=6.25cm]{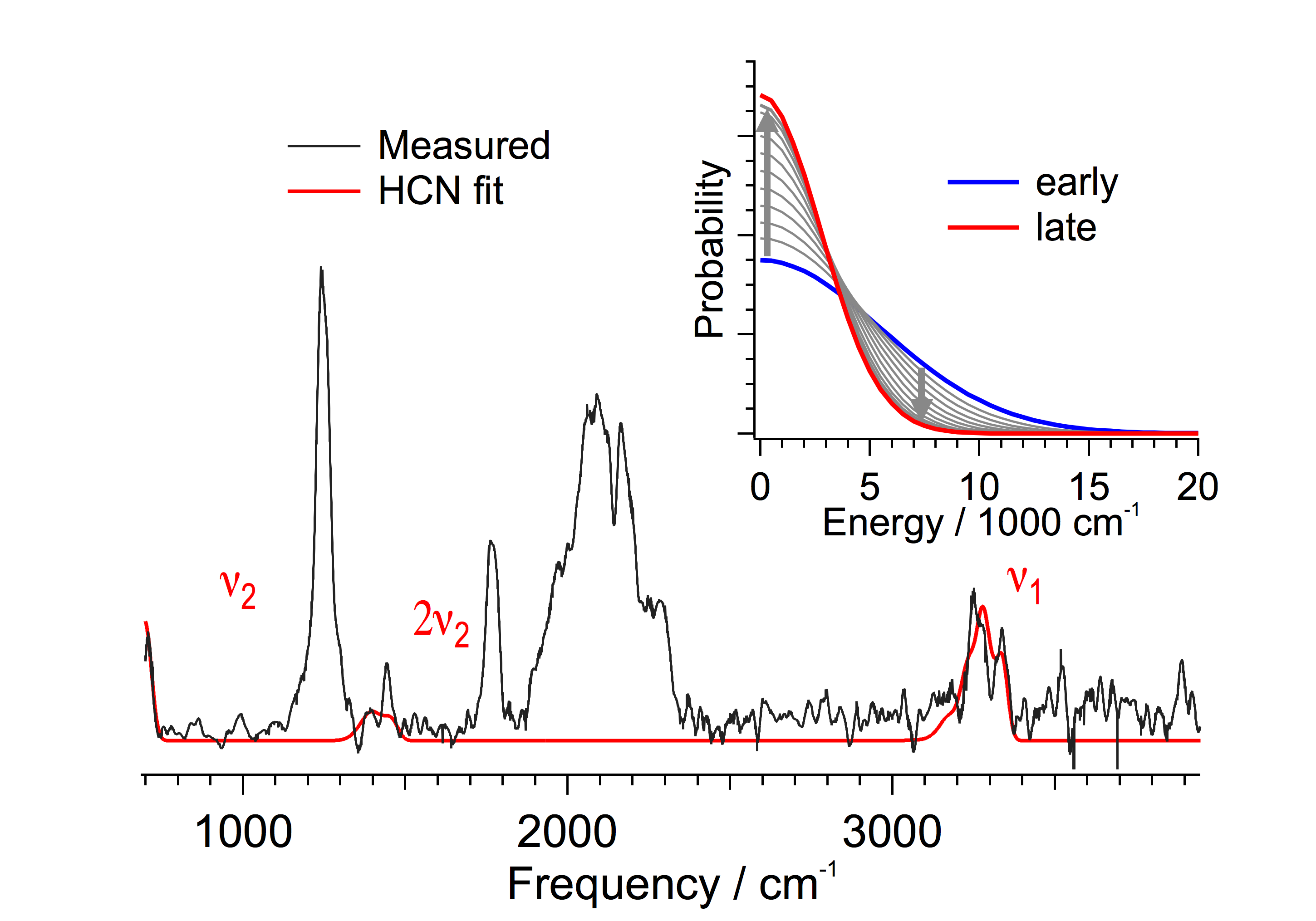}
  \caption{Representative best fit of HCN emission resonances (red curve) in a late-time (12 $\mu$s) reconstructed IR emission spectrum (black curve). Inset depicts fit-deduced time-resolved HCN probability distributions as a function of HCN internal energy.}
\end{figure}
\subsection{\bf{\em{NCCO + CH$_3$O dissociation channel.}}}
The most likely electronic transitions occurring upon 193 nm photon absorption lead to excited singlet states, because of spin
\begin{figure}[h]
 \centering
  \includegraphics[height=6.25cm]{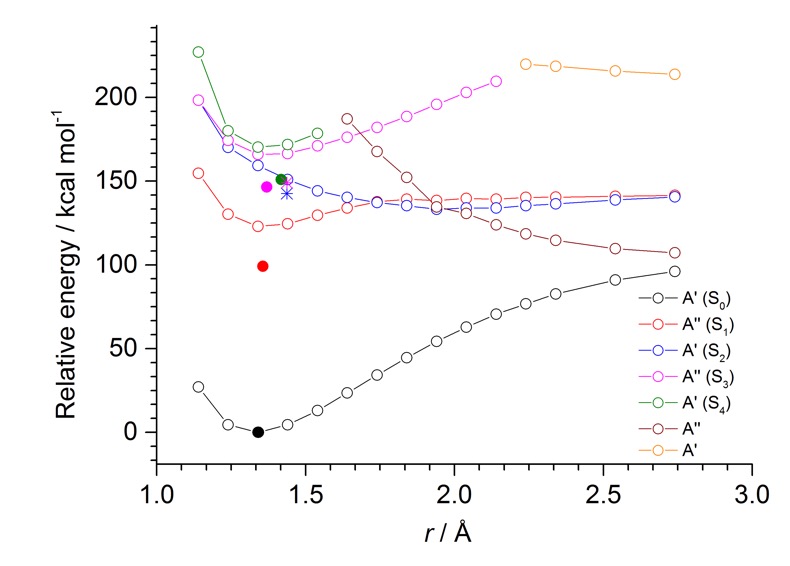}
  \caption{Potential energy curves calculated at the MS-CASPT2//SA-CASSCF level for the lowest lying singlet states. The energies are relative to the minimum energy of the S$_0$ curve. The solid circles correspond to equilibrium structures. The stars are the MS-CASPT2 energies of states S$_2$ (blue) and S$_3$ (magenta) for a minimum energy crossing point between these two surfaces obtained at the SA5-CASSCF.}
\end{figure}
conservation and the fact that the ground electronic state of MCF is a singlet. Therefore, in this work, we only paid attention to the lowest lying singlet states. Moreover, since there are no heavy atoms involved in the reaction, we can neglect spin-orbit coupling.\\
\indent To study the formation of NCCO + CH$_3$O, we calculated potential energy curves for this dissociation in the ground electronic state and four lowest singlet excited states. As shown below, this number of states is enough to investigate the excitation corresponding to a photon wavelength of 193 nm.\\
\indent The calculated potential energy curves are depicted in Figure 5. The open circles in the figure correspond to a series of ground state geometry optimizations at the SA5-CASSCF(12,10)/cc-pVTZ level, constraining the O-C distance. Final energies were obtained by single-point MS5-CASPT2/cc-pVTZ calculations. The solid circles show the MS5-CASPT2 energies calculated for the equilibrium SA5-CASSCF structures in the different electronic states.\\
\indent The geometries and the energies of all these structures are collected in the SI. All the full and constrained optimizations led to geometries showing nearly C$_S$ symmetry, and therefore the electronic states may be classified according to the corresponding symmetry species ($A'$ or $A''$).\\
\indent One of the most relevant results shown graphically in Figure 5 is the presence of a repulsive state, the S$_2$ state, depicted in blue in the figure. This state has a $\pi\pi^*$ nature, as shown in Table 1S (see SI), which gives the main configurations for the calculated excited states. More details about the nature of the excited states investigated here are presented in Figure 2S in the SI, which describes the differences between the electron densities of the excited states and that of the ground state. The question now is, {\em{Does this electronic state play a fundamental role in the 193 nm photodissociation of MCF?}} This is readily answered by inspection of Figure 6, which shows the simulated UV absorption spectrum of MCF, considering the contributions from excited states S$_1$ to S$_4$.
\begin{figure}[h]
 \centering
  \includegraphics[height=6.25cm]{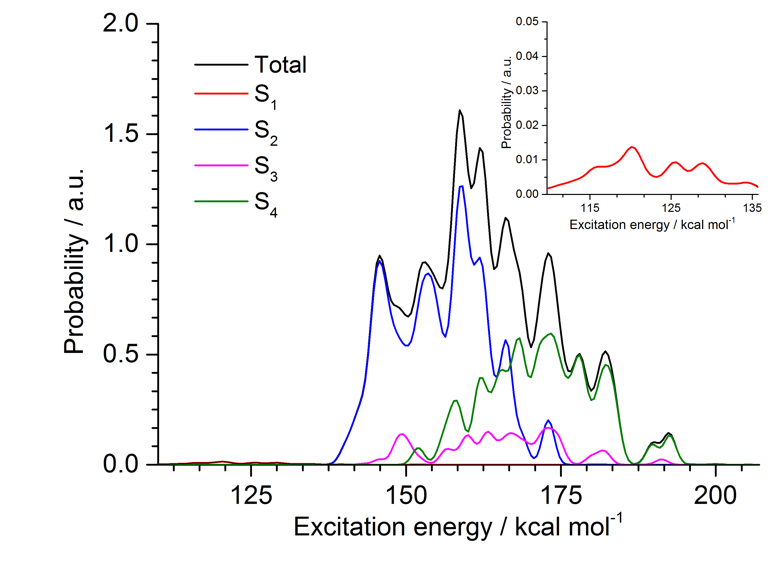}
  \caption{Simulated absorption spectrum of MCF. The total spectrum (black line) is composed of contributions from transitions to S$_1$ (red line), S$_2$ (blue line), S$_3$ (magenta line) and S$_4$ (green line). The inset displays an expanded view of the spectrum for the 110-136 kcal/mol region, wherein excitation leads to the S$_1$ state only.}
\end{figure}\\
\indent According to this simulated spectrum, at 193 nm (148.1 kcal/mol) MCF is mostly excited to the S$_2$ state, as this state (blue line) contributes 87.4\% to the total absorption. Therefore, our calculations point out that NCCO + CH$_3$O formation is the dominant decomposition channel in the photodissociation of MCF at 193 nm, in agreement with the experimental observations, and that this process takes place on an excited-state potential energy surface. In addition, we notice that the S$_2$ state correlates with CH$_3$O and NCCO products in the ground and excited states, respectively, which is consistent with our observation of electronic excitation of the CN photofragment.\\
\indent On the other hand, the simulated spectrum indicates that at 193 nm, 12.6\% of the electronic transitions go to the S$_3$ state (magenta line). This is a bound state with an equilibrium structure having a C-O bond length of 1.370 $\AA$ (solid magenta circle). Importantly, we found a minimum energy crossing point between the S$_2$ and S$_3$ surfaces, which may be easily accessible after a transition to the S$_3$ state because it is located, both geometrically and energetically, close to those of the S$_3$ equilibrium structure (see Figure 5). Notice that the crossing point in Figure 5 is represented by two points (blue and magenta stars) due to differences between the CASSCF and CASPT2 potential energy surfaces. This S$_2$/S$_3$ crossing point opens another pathway for dissociation.\\
\indent Although our calculations support that the majority of the excited MCF molecules will exhibit NCCO + CH$_3$O dissociation on an excited potential energy surface, they also open the possibility of trapping population in the S$_1$ state, since there is a minimum lying well below the dissociation limit and clear evidence of degeneration between S$_1$ and S$_2$. This could allow a fraction of the molecules to return to the ground electronic state via internal conversion, and proceed along any of the ground state dissociation channels in ground state reaction{\color{blue} \cite{Rubio-Lago2012}}.
\subsection{\bf{\em{Dissociation channels in the ground electronic state.}}}
The use of our automated TS search procedure results in the finding of an important number of minima, TSs, and dissociation channels in the S$_0$ state of MCF. Their relative energies and geometries are gathered in the SI.\\
\indent In particular, the three most stable isomers of the system are acetyl isocyanate, cyanoacetic acid, and oxazolone, whose relative energies (with respect to MCF) are: -20.6, -16.9 and -5.7 kcal/mol, respectively. However, no pathways connecting these isomers with MCF have been found, and therefore they were not included in the kinetic study. In the following, only the most relevant structures, in terms of their importance in the 193 nm photolysis of MCF, are discussed.\\
\indent Figure 7 shows the most important dissociation channels of MCF in the S$_0$ state (pathways with yields $<$1\% are not included for simplicity). MCF has two conformational isomers (MIN 4 and MIN 9), connected by a transition state (TS 144) lying 11.6 kcal/mol over MCF. The higher energy conformational isomer MIN 9 is indeed the reactive one, as seen in the figure. MIN 4 and MIN 9 are connected with the corresponding conformers of methyl isocyanoformate (MisoCF), MIN 17 and MIN 20, via transition states TS 12 and TS 13, respectively. These pathways present barrier heights of approximately 50 kcal/mol. Additionally, the two isomers of MisoCF are also connected via a low energy transition state (TS 145).
\begin{figure}[t]
 \centering
  \includegraphics[height=6.75cm]{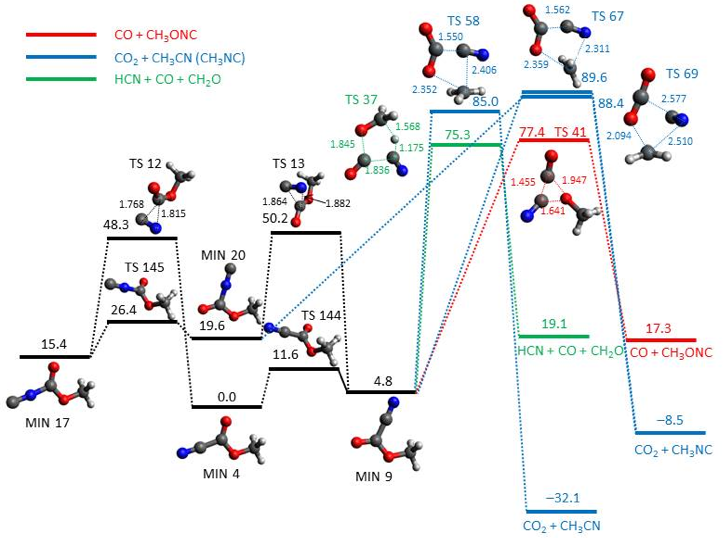}
  \caption{Most important (yields greater than 1\%, see the SI) dissociation channels of MCF at 148.12 kcal/mol in the ground electronic state. The different colors refer to: isomerization (black), CO elimination (red), CO$_2$ elimination (blue), and the triple channel leading to HCN + CO + CH$_2$O (green). The numbers are relative energies in kcal/mol (including the zero-point vibrational energy) with respect to MCF, calculated at the CCSD(T)/6-311++G(3df,3pd)//MP2/6-311+G(2d,2p) level of theory with the vibrational frequencies obtained using MP2/6-311+G(2d,2p). The dotted lines in the structures indicate the bonds that are being broken/formed.}
\end{figure}\\
\indent As shown in Figure 7, the three major dissociation channels in the S$_0$ state lead to the following products: 1) CO + C$_2$H$_3$NO, 2) CO$_2$ + C$_2$H$_3$N, and 3) HCN + CH$_2$O + CO. The kinetic simulations yield relative abundances of 47\%, 40\% and 13\%, respectively. CO dissociation is the predominant channel (red in Figure 7), and involves the three-center transition state TS 41, whose relative energy is 77.4 kcal/mol. The second most important channel is CO$_2$ elimination (blue in Figure 7), which can be obtained via three different paths. Two of them lead to methyl isocyanide as co-product, while the third one leads to methyl cyanide. The CO$_2$ + CH$_3$CN channel is obtained from MIN 9, via transition state TS 58 (relative energy of 85 kcal/mol). From the same minimum departs another path leading to CO$_2$ + CH$_3$NC, via the higher-energy (89.6 kcal/mol) transition state TS 67. Finally, the same products (CO$_2$ + CH$_3$NC) can be reached from MIN 20, through TS 69, whose relative energy is 88.4 kcal/mol.\\
\indent Finally, the most important HCN dissociation channel involves a triple (or three-body) path via transition state TS 37 (green in Figure 7) leading to HCN + CO + CH$_2$O. The corresponding triple dissociation channel leading to HNC + CO + CH$_2$O has also been found within this study (see the SI), but its yield is lower than 1\%.\\
\indent The kinetic results carried out in this work in the ground electronic state suggest that, so long as a fraction of molecules undergo cascading internal conversion from nascent S$_2$ population to the S$_0$ state as our calculations suggest, HCN (and to a smaller extent HNC) should be observed in the 193 nm photolysis of MCF. The experimental evidence we provide in this study for HCN formation from photolysis of MCF is corroborated by the finding of the HCN dissociation channel depicted in Figure 7.
\section{DISCUSSION}
\subsection{\bf{\em{Photodissociation dynamics of MCF at 193 nm.}}}
Furlan et al. previously used PTS to characterize the photolysis of MCF at 193 nm{\color{blue} \cite{Furlan2000}}. MCF is a cyano-ester and was predicted to follow Norrish type I reactivity. It was deduced that the photolysis favors production of the NCCO radical (reaction 1, $\Phi$=95\%), though a non-negligible radical CN channel was also detected (reaction 2, $\Phi$=5\%). Given our present observation of the unexpected primary dissociation channels yielding HCN and HNC, a re-examination of the 193 nm photolysis of MCF is in order.\\
\indent Firstly, it is of interest to question, {\em{Why was the production of HCN and HNC overlooked in the PTS study?}} The simple answer is that there was no {\em{a priori}} reason to search for these photofragments. HCN cannot be produced directly from MCF via a pure Norrish type I dissociation. Further, strictly speaking, due to the absence of a hydrogen located gamma to the carbonyl, a pure Norrish type II reaction is likewise not possible for MCF. Ultimately, in a study aimed at characterizing Norrish type reactivity, there simply was no incentive to look for signal assignable to either HCN or HNC. Nevertheless, just as in the case of Lee et al.Õs investigation of pyrimidine{\color{blue} \cite{Lin2006}}, the PTS results alone would have been insufficient to distinguish HCN from HNC.\\
\indent Furthermore, despite the apparent prominence of the HCN features in the reconstructed emission spectra (Figure 1b), production of HCN is actually a relatively minor elimination channel for this system. It is only because emission spectra selectively bias vibrationally excited species, as well as the comparatively strong transition dipole moments of HCN, that we observe any HCN at all. Indeed, prior to application of SRa to improve the SNR, the HCN transitions were completely obscured by the background noise of the spectra (Figure 1a).\\
\indent Of significance, the fit analysis of the nascent emission spectra reveal that, relative to the minor radical CN channel, HCN is produced with even lower quantum efficiency. Specifically, the experimentally deduced nascent populations of ca. 6.4$\times$10$^{-8}$ (HCN) and 7.7$\times$10$^{-7}$ (CN) suggest relative branching ratios of 92.3\% and 7.7\% for the CN and HCN channels, respectively. Further, recall that Furlan et al. measured the branching ratio for the CN elimination channel as ca. 5$\pm$2\%{\color{blue} \cite{Furlan2000}}. Subsequently, using CN as an internal reference, it is possible to rescale the relative branching ratio and estimate the absolute branching ratio of the HCN elimination channel. Specifically, assuming that all nascent CN is produced as electronically excited CN(A$^2\Pi_1$), a lower-bound limit of the absolute branching ratio for the HCN elimination channel can be deduced as 0.42$\pm$0.24\%. Unfortunately, due to the limited available signal, a robust spectral fit analysis of the HNC band is not feasible with this data. Nevertheless, by comparing the integrated emission intensity of the HCN and HNC bands in the late time spectra, it is possible to estimate the relative HNC/HCN branching ratio as ca. 0.1 (see SI for details).\\
\indent We note that the experimentally deduced HCN branching ratio is well supported by the calculated excited state potential energy curves (Figure 5) and corresponding simulated UV absorption spectra (Figure 6). Specifically, excitation at 193 nm is predicted to result in transitions to the repulsive S$_2$ state, predominantly yielding dissociation along the Norrish type I elimination channel to NCCO + CH$_3$O (reaction 1). More importantly, the calculated equilibrium minimum energy of the bound S$_1$ state falls below the S$_0$ ground state dissociation limit. Subsequently, due to degeneracies between S$_2$ and S$_1$, it is possible (though not probable) to internally convert nascent S$_2$ population to S$_1$, and then to vibrationally highly excited levels of S$_0$, where the ground state dissociation channels are invoked. Even then, however, RRKM and KMC kinetic simulations predict that dissociation along the HCN elimination channel only accounts for ca. 13\% of the (comparatively minor) ground state branching ratio. Worse yet, HNC is predicted to account for less than 1\% of the ground state product yield. Overall, these predicted branching ratios are in semi-quantitative agreement with the experimental observation that the HCN and HNC elimination channels are comparatively minor (ca. 0.4\% and 0.04\%, respectively). Subsequently, even if HCN or HNC had been expected photoproducts, it is likely that the corresponding signals would have been undetectable in the PTS study.\\
\indent Next, it is of interest to examine the mechanisms leading to the production of HCN and HNC. Again, due to the molecular structure, a pure Norris type II dissociation is not feasible for MCF. Nevertheless, it is worth noting that the three methoxy hydrogens are all gamma to the cyano group, and could conceivably yield HCN (or HNC) via a cyano-analogue of the Norrish type II reaction. Specifically, the in-plane methoxy hydrogen could be abstracted by the cyano group, resulting in the transient production of a radical intermediate. Subsequent rearrangement would result in bond cleavage, ultimately yielding: HCN, CO, and CH$_2$O. As depicted in Figure 7 (green path), this is fully consistent with the predicted HCN channel deduced using the TSSCDS automated TS search algorithm. However, rather than a sequential abstraction/cleavage reaction, the TSSCDS predicted TS corresponds to a concerted three-body dissociation process. A similar pathway is predicted for the production of HNC, though involves a preliminary isomerization to the corresponding isocyanide, MisoCF. The validity of the proposed elimination mechanism is revealed through characterization of the nascent HCN population distribution.  Specifically, Figure 8 depicts a direct comparison of the measured and QCT simulated nascent population distribution of HCN. Of significance, the observed and predicted population distributions are in near perfect agreement.
\begin{figure}[t]
 \centering
  \includegraphics[height=6.cm]{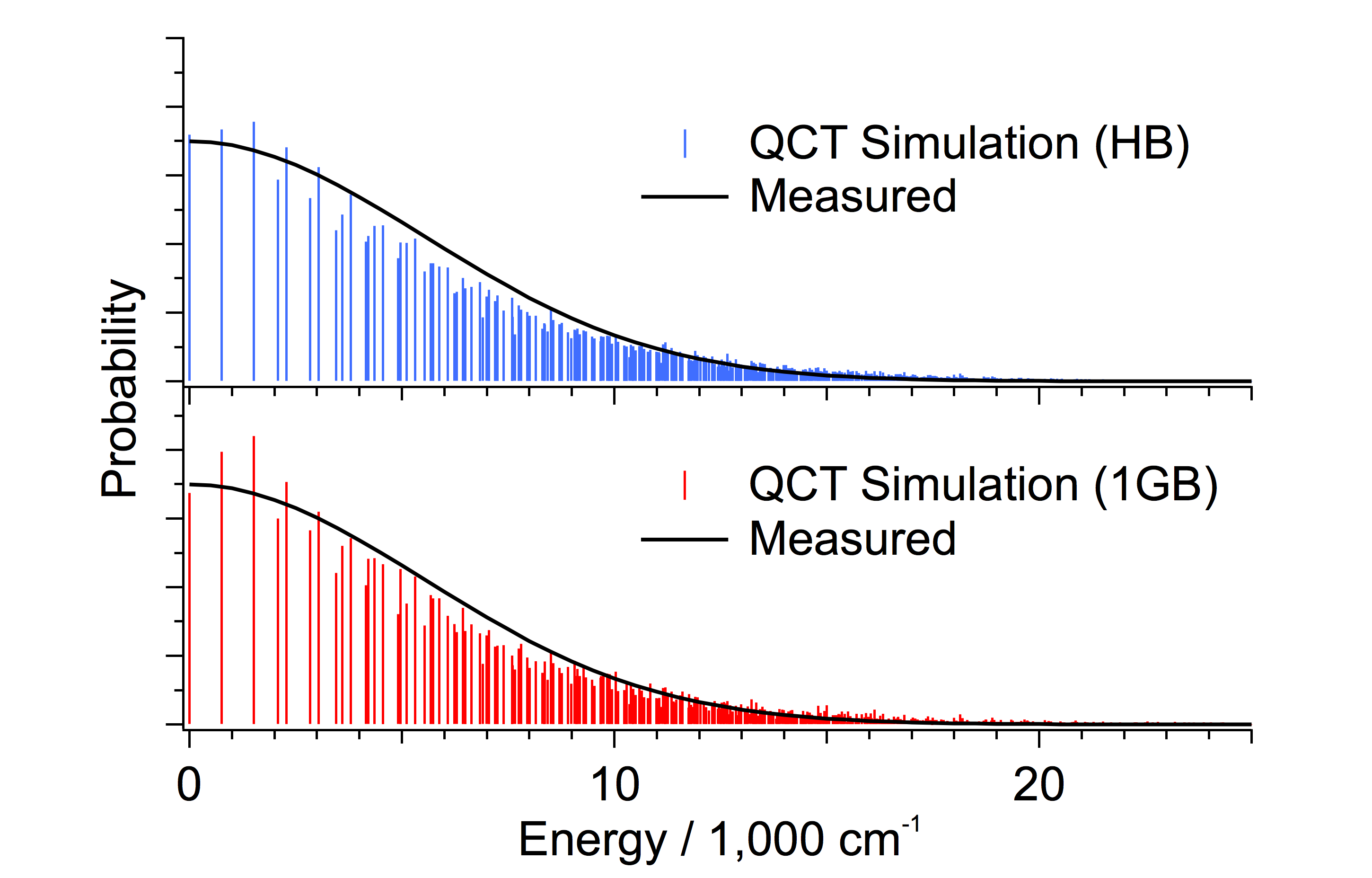}
  \caption{Direct comparison of the measured (black line) and QCT simulated (top blue, bottom red lines) nascent population distribution of HCN.}
\end{figure}\\
\indent Finally, in addition to MCF, Furlan et al. also characterized the 193 nm photolysis of acetyl cyanide (NCC(O)CH$_3$){\color{blue} \cite{Furlan2000}}. The acetyl radical is notoriously unstable, and readily dissociates to CO + CH$_3$ (ca. 6000 cm$^{-1}$ dissociation energy){\color{blue} \cite{Furlan2000}}. Subsequently, due to the observation of stable acetyl radical (and conservation of energy arguments), it was predicted that either the acetyl or CN radical was produced electronically excited, though it was unclear which was the case. This is fully consistent with our observation of nascent electronically excited CN(A$^2\Pi_1$), which is likely produced from primary photolysis via a Norrish type I dissociation (reaction 2). This was also observed in the prior study of the 193 nm photolysis of the symmetric ketone, carbonyl cyanide (NCC(O)CN){\color{blue} \cite{McNavage2004}}. Additionally, it is of interest to note that dissociation of MCF from the repulsive S$_2$ state correlates with CH$_3$O and NCCO products in the ground and excited states, respectively. This suggests another viable source for electronically excited CN(A$^2\Pi_1$). However, unless the secondary dissociation of electronically excited NCCO is prompt, and therefore comparatively rapid on the timescale of our detection, this source likely does not contribute significantly to the nascent CN(A$^2\Pi_1$) signal in Figure 3.\\
\\
{\bf{\em{UV photolysis as a source of astrophysical HCN / HNC.}}}\vspace{0.5em}\\
We now return to our original assertion that cyano-containing hydrocarbons are viable photolytic sources for HCN and HNC, and therefore potentially useful for explaining the observed overabundance of astrophysical HNC. As discussed above, this is certainly true of MCF, in which production of both cyanides were observed following 193 nm photolysis. Despite the fact that the combined quantum yields for both HCN and HNC account for less than 1\% of the total product yield, this does not diminish the significance of the observation. The magnitude of the quantum yield is largely a result of our chosen photolysis wavelength. Indeed, as shown in Figure 6 and discussed above, excitation of MCF at 193 nm primarily results in transitions to the repulsive S$_2$ state, which favors elimination along the Norrish type I dissociation channel. However, at lower energy wavelengths, for instance near 236 nm (121.1 kcal/mol), the only possible transition is to the bound S$_1$ state (see Figure 6). This would lead to a significant enhancement in the quantum yield from the ground-state dissociation channels. To our knowledge, MCF has not yet been observed in the ISM. Regardless, even at longer wavelengths, UV photolysis of MCF is unlikely to be a dominant source of astrophysical HCN or HNC. Nevertheless, the core value of the current result is the demonstration that cyano-esters (and likely cyano-ketones also) are viable photolytically sources of HCN and HNC via a mechanism analogous to the Norrish type II reaction.\\
\indent Beyond the simple photolytic production of the cyanides, it is critical to note that, due to conservation of energy, the cyanides are produced ro-vibrationally hot. As depicted in Figure 8, MCF produces nascent HCN with in excess of 28.6 kcal/mol (10,000 cm$^{-1}$) of internal energy. This is comparable to our prior characterization of the 193 nm photolysis of vinyl cyanide, in which the average nascent internal energies of HCN and HNC were deduced to be 25.2 kcal/mol and 24.5 kcal/mol, respectively{\color{blue} \cite{Wilhelm2009}}. In addition to increasing the propensity for isomerization, the high internal excitation of the cyanides invariably adds to their reactivity, and potentially to the anomalous over production of astrophysical HNC.
\section{CONCLUSIONS}
We report the observation of new primary dissociation channels in the 193 nm photolysis of MCF. Specifically, application of spectral reconstruction analysis{\color{blue} \cite{Wilhelm2015d}} to improve the SNR (ca. 5x enhancement) of measured time-resolved FTIR emission spectra revealed the presence of previously obscured resonances assignable to ro-vibrationally excited HCN and HNC. Spectral fit analysis of the nascent HCN and CN(A$^2\Pi_1$) resonances allowed direct determination of the relative populations of each photofragment. Subsequent comparison against the previously deduced CN branching ratio (ca. 5\%) suggest a total HCN branching ratio of 0.42$\pm$0.24\%. Conversely, signal from HNC was too weak for a robust fit analysis, but a ratio of the measured integrated emission intensity allowed an estimate of 0.1 for the HNC/HCN relative abundance ratio.\\
\indent Multi-configuration self-consistent field calculations were used to characterize the ground and four lowest energy singlet excited states of MCF. Excitation at 193 nm primarily promotes transitions to the repulsive S$_2$ state, resulting in the production of NCCO + CH$_3$O via a Norrish type I reaction. Nevertheless, it was deduced that cascading internal conversion from nascently prepared S$_2$ state MCF could produce minor populations of vibrationally highly excited ground state MCF. Application of the recently developed automated TS search algorithm{\color{blue} \cite{Martinez-Nunez2015, Martinez-Nunez2015a}} revealed the presence of discrete ground state dissociation channels producing HCN and HNC, which were consistent with a cyano-analogue of the Norrish type II reaction. The deduced mechanism was validated by direct comparison of the QCT predicted and experimentally measured nascent HCN population distributions, which were found to be in near quantitative agreement. Consistent with the experimentally deduced branching ratios, RRKM and KMC simulations predict that, of the ground state dissociation channels, production of HCN and HNC occurs with a quantum yield of 13\% and $<$1\%, respectively.\\
\indent Finally, we now show that cyano-esters, such as MCF, are viable photolytic sources of HCN and HNC. Furthermore, similar to prior studies of vinyl cyanide{\color{blue}\cite{Wilhelm2009, Homayoon2011, Vazquez2015}}, photolytic production yields HCN and HNC products that are ro-vibrationally hot; thereby enhancing their prospects for subsequent reactivity (e.g., isomerization, reaction), and potentially aiding in the anomalous over production of astrophysical HNC.
\section{ACKNOWLEDGEMENTS}
This work was supported in part through the U.S. Department of Energy, Office of Basic Energy Sciences, Grant No. DEFG 02-86ER 134584, and by Xunta de Galicia through project GRC2014/032. The authors also thank {\em{Centro de Supercomputaci\'{o}n de Galicia (CESGA)}} for the use of their computational facilities.

\bibliography{mcf_bib} 
\bibliographystyle{phjcp} 

\end{document}